\documentclass[aps,preprintnumbers,showpacs,twocolumn,superscriptaddress,floatfix,nofootinbib]{revtex4-1}

\usepackage{latexsym} 
\usepackage{amssymb} 
\usepackage{amsfonts}
\usepackage{amsmath} 
\usepackage{bm} 
\usepackage{mathtools}
\usepackage{envmath}
\usepackage{color}
\usepackage{times} 
\usepackage{hyperref}
\usepackage{float}
\usepackage{latexsym}
\usepackage{envmath}

\def\prl{Phys. Rev. Lett.}
\def\prd{Phys. Rev. D}
\def\cqg{Class. Quantum Grav.}

\def\length{\sigma}
\def\Dflat{{\mathcal D}}

\usepackage{amssymb,amsmath}
\usepackage{graphicx} 

\begin{document}
\title{Critical phenomena in the gravitational collapse of electromagnetic waves}

\author{Thomas W. \surname{Baumgarte}}\affiliation{Department of Physics and Astronomy, Bowdoin College, Brunswick, ME 04011, USA}

\author{Carsten \surname{Gundlach}}\affiliation{Mathematical Sciences, University of Southampton, Southampton SO17 1BJ, United Kingdom}

\author{David \surname{Hilditch}}\affiliation{CENTRA, Departamento de F\'isica, Instituto Superior T\'ecnico IST, Universidade de Lisboa UL, Avenida Rovisco Pais 1, 1049 Lisboa, Portugal}


\begin{abstract}
We numerically investigate the threshold of black-hole formation in
the gravitational collapse of electromagnetic waves in axisymmetry.  We
find approximate power-law scaling $\rho_{\rm max}\sim (\eta_*-\eta)^{-2\gamma}$  of the maximum density in the time evolution of
near-subcritical data with
$\gamma\simeq 0.145$, where $\eta$ is the amplitude of the initial data. We
directly observe approximate discrete self-similarity in near-critical
time evolutions with a log-scale echoing period of $\Delta\simeq 0.55$. The
critical solution is approximately the same for two families of
initial data, providing some evidence of universality.  Neither the discrete
self-similarity nor the universality, however, are exact.  We speculate that the absence
of an exactly discrete self-similarity might be caused by the interplay of electromagnetic and
gravitational wave degrees of freedom, or by the presence of higher-order angular multipoles, or both, and discuss
implications of our findings for the critical collapse of vacuum gravitational waves.
\end{abstract}


\maketitle

Critical phenomena in gravitational collapse were first reported in the seminal work of Choptuik \cite{Cho93}, who considered families of initial data 
parameterized by a parameter $\eta$, say.  In dynamical evolutions, subcritical data with sufficiently small  $\eta$ disperse to infinity, leaving behind flat space, while supercritical data with sufficiently large $\eta$ form a black hole.  Critical phenomena can then be observed in the vicinity of the critical parameter $\eta_*$ that separates the two regimes.   Specifically, evolutions close to criticality approach a self-similar critical solution, and dimensional quantities characterizing the evolution follow a power-law.  The mass of a black hole formed in supercritical evolutions, for example, will satisfy
\begin{equation} \label{mass_scaling}
M \simeq (\eta - \eta_*)^\gamma,
\end{equation}
while the maximum density observed during subcritical evolutions scales with 
\begin{equation} \label{rho_scaling}
\rho_{\rm max} \simeq (\eta_* - \eta)^{-2 \gamma},
\end{equation}
where $\gamma$ is the critical exponent (see also \cite{GarD98}).

Many aspects of these phenomena are quite well understood, at least in the context of spherical symmetry (see, e.g., \cite{Gun03,GunM07} for reviews).  In many cases for which there exists a spherically symmetric critical solution, there is compelling numerical evidence that this critical solution is unique (for a given matter model) and is either discretely self-similar (DSS, for example for massless scalar fields \cite{Cho93}) or continuously self-similar (CSS, for example for radiation fluids \cite{EvaC94}).   The critical exponent $\gamma$ is the inverse of the Lyapunov exponent of linear perturbations of the critical solution \cite{KoiHA95,Mai96,Gun97}.  For critical solutions that are DSS, the oscillations in the critical solutions are reflected by a periodic ``wiggle" that is superimposed on the scaling laws (\ref{mass_scaling}) and (\ref{rho_scaling}) (see \cite{Gun97,HodP97}).  

The situation is much less clear when a critical solution -- should it exist -- cannot be expected to be spherically symmetric.  Perhaps the most important example is the gravitational collapse of (vacuum) gravitational waves.   While critical phenomena in this collapse were first reported in the pioneering work of Abrahams and Evans \cite{AbrE93,AbrE94}, their results have yet to be confirmed independently, despite the attempts of a number of groups and researchers (see, e.g., \cite{AlcABLSST00,GarD01,San06,Rin08,Sor11,HilBWDBMM13}).    Numerical work seems to be hampered by the difficulty of finding suitable coordinate conditions (see also \cite{HilWB17,KhiL18}), while analytical or semi-analytical approaches are quite  complicated because of the inherent absence of spherical symmetry (compare with \cite{Gun95,MarG03} for similar constructions for scalar fields in spherical symmetry).  Recent progress by \cite{HilWB17} has confirmed the critical exponent of $\gamma \simeq 0.37$ reported by \cite{AbrE93}, but has also raised new questions about the nature of the critical solution.

Motivated by these considerations we study in this paper critical phenomena in the gravitational collapse of electromagnetic waves, which, to the best of our knowledge, have not been considered before.  We focus on axisymmetry, in which case Maxwell's equations can be reduced to a single wave equation that, in many ways, is similar to that for a scalar field.  On the other hand, our setup shares with the vacuum gravitational wave case the property that a critical solution cannot be spherically symmetric.  In this sense we hope that our work will also help our understanding of the critical collapse of gravitational waves.  

Axisymmetry is generated by a Killing vector field that, in adapted coordinates, takes the form $\xi = \partial /\partial \varphi$.  We also assume equatorial symmetry (the reflection $\theta \to \pi - \theta$), which, together with axisymmetry, singles out the worldline of a preferred central observer.

We express Maxwell's equations in terms of a vector potential $A_a$, so that the Faraday tensor is $F_{ab} = \nabla_a A_b - \nabla_b A_a$.  Here $\nabla_a$ is the covariant derivative associated with the spacetime metric.  We employ a 3+1 foliation of the spacetime, and introduce $n^a = \alpha^{-1} ( 1 , - \beta^i)$ as the normal vector on spatial slices, where $\alpha$ is the lapse function and $\beta^i$ the shift vector.   Without loss of generality we can choose a gauge with $\Phi \equiv n^a A_a = 0$, so that $A_a$ becomes purely spatial.  In the absence of charges, Maxwell's equations can then be written as
\begin{subequations} \label{maxwell}
\begin{eqnarray} 
d_t A_i & = & - \alpha E_i \\
d_t E^i & = & - D_j (\alpha D^j A^i) + D_j (\alpha D^i A^j) + \alpha K E^i
\end{eqnarray}
\end{subequations}
together with the constraint $D_i E^i = 0$.  Here $E^i$ is the electric field, $D_i$ the covariant derivative associated with the spatial metric $\gamma_{ij} \equiv g_{ij} + n_i n_j$, $K = - \nabla_a n^a$ the mean curvature, and $d_t \equiv \partial_t - {\mathcal L}_\beta$.  In twist-free axisymmetry (see \cite{Ger71}), Maxwell's equations (\ref{maxwell}) can be reduced to a single wave equation for $A_\varphi \equiv \xi^a A_a$, and the stress-energy tensor $T_{ab}$ can be computed from $E^\varphi$ and spatial derivatives of $A_\varphi$.\footnote{In axisymmetry with a twist, the electromagnetic field and its stress-energy tensor can be expressed in terms of $A_\varphi$ and a new field $\tilde A_\varphi$, where the potential $\tilde A_b$ generates the dual ${}^*F_{ab}$ of the Faraday tensor. Both $A_\varphi$ and $\tilde A_{\varphi}$ are again gauge-invariant.}   In particular, the energy density as observed by a normal observer is 
\begin{equation} \label{density}
\rho \equiv n_a n_b T^{ab} = \frac{1}{8\pi} \left( E_i E^i + B_i B^i \right),
\end{equation}
where $B^i = \epsilon^{ijk} D_j A_k$ is the magnetic field.

We choose time-symmetric and conformally flat initial data, so that $\gamma_{ij} = \psi^4 \hat \gamma_{ij}$ where $\psi$ is a conformal factor and $\hat \gamma_{ij}$ the flat metric.  At $t = 0$ we further choose $A_\varphi = 0$ and
\begin{equation} \label{E_initial}
E^\varphi = - \frac{4 \eta}{\psi^6  \length^2} \left( e^{- (r - r_0)^2/\sigma^2} + e^{ - (r + r_0)^2/\sigma^2} \right),
\end{equation}
which satisfies the constraint $D_i E^i = 0$ identically (we note that either $A_i = 0$ or $E^i = 0$ is consistent with time-symmetric initial data, since in either case $S_i \equiv \gamma_{ia} n_b T^{ab} = 0$).  Here $\eta$ is a dimensionless amplitude, $r$ is our radial coordinate, $r_0$ determines the location of the maximum of $E^\varphi$, and $\sigma$ is a constant of unit length.  In the following we will present all dimensional quantities in units of $\sigma$, which is equivalent to setting $\sigma = 1$ in the above.   Given a guess for $\psi$, we compute the density $\rho$ from (\ref{E_initial}) and (\ref{density}), and then solve the Hamiltonian constraint
\begin{equation} \label{Hamiltonian}
\Dflat^2 \psi = \hat \gamma^{ij} \Dflat_i \Dflat_j \psi = - 2 \pi \psi^5 \rho,
\end{equation}
where $\Dflat_i$ is the covariant derivative associated with the flat metric $\hat \gamma_{ij}$, to recompute $\psi$, iterating until convergence to within a given tolerance has been achieved.  We inserted the factor of $\psi^6$ in (\ref{E_initial}) in order to make the solution to (\ref{Hamiltonian}) unique; see, e.g.,~\cite{Yor79,BauMP07}.

We then solve the Maxwell-Einstein system by evolving the electromagnetic fields according to Maxwell's equations (\ref{maxwell}) together with Einstein's equations for the gravitational fields.  We adopt the BSSN formulation of Einstein's equations \cite{NakOK87,ShiN95,BauS98}, 
implemented in spherical polar coordinates $r$, $\theta$ and $\varphi$ \cite{MonC12,BauMCM13,BauMM15} with the help of a reference-metric formulation (see, e.g., \cite{BonGGN04,ShiUF04,Bro09,Gou12}).   We also rescale components of tensorial quantities with factors of $r$ and $\sin \theta$, so that singular terms at the center and on the axis can be handled analytically.  Specifically, we evolve the functions $a_\varphi \equiv A_\varphi / (r \sin \theta)$ and 
$e^\varphi \equiv E^\varphi r \sin \theta$ rather than $A_\varphi$ and $E^\varphi$ themselves.   We use a grid setup similar to those used in the critical collapse simulations of \cite{BauM15,BauG16,GunB16,GunB18,CelB18,Bau18}, except that we implement an asymptotically logarithmic grid using the approach of \cite{RucEB18}, allowing the innermost radial grid-cell to be about $4 \times 10^{-3}$ the size of the outermost grid-cell, and use the method of lines with a fourth-order Runge Kutta method for the time evolution rather than the PIRK method \cite{MonC12}.  As in the above references we allow for radial regridding during the evolution, and start with the outer boundary at $r_{\rm out} = 128$.  Unless noted otherwise we show results for $N_r = 192$ radial and $N_\theta = 18$ angular grid-points. 

We evolve the initial data with the ``one plus log" slicing condition \cite{BonMSS95} with a ``pre-collapsed" lapse $\alpha = \psi^{-2}$ as initial data.  Similar to experience with simulations of vacuum gravitational waves, we found that ``Gamma-freezing" shift conditions \cite{AlcBDKPST03,ThiBB11} do not lead to stable evolutions in the vicinity of the black-hole threshold (compare \cite{HilBWDBMM13,KhiL18}).  We instead choose zero shift.  While this choice leads to instabilities once black holes form in supercritical evolutions, it allowed us to approach the threshold with subcritical data.

In the following we consider two families of the initial data (\ref{E_initial}), a ``centered" family with $r_0 = 0$, and an ``off-centered" family with $r_0 = 3$.  For both families we fine-tune the parameter $\eta$ up to about $|\eta - \eta_*| \simeq 10^{-11}$ of the threshold parameter $\eta_*$, which we find to be $\eta_* \simeq 0.913$ for $r_0 = 0$ and $\eta_* \simeq 0.0703$ for $r_0 = 3$.  We then analyze the density $\rho$ at the center (where it takes an invariant meaning), as well as 
\begin{equation} \label{A_xi}
A_\xi \equiv \frac{\xi^a A_a}{\sqrt{\xi^a \xi_a}} = \frac{A_\varphi}{\sqrt{g_{\varphi\varphi}}}
\end{equation}
in order to probe the properties of the critical solution.\footnote{We will refer to the solution at the threshold of black-hole formation as the critical solution, even though we cannot establish that this solution is unique.}

A spacetime is DSS if there exists a discrete conformal isometry $\Phi$ such that $\Phi^*g_{ab}=e^{-2\Delta} g_{ab}$, that is, the spacetime looks the same when all proper lengths and times have been shrunk by a factor of $e^{-\Delta}$.  Matter fields scale consistently with the Einstein equations.  In our system, this means $\Phi^*A_\xi=A_\xi$.  In order to analyze this behavior we introduce auxiliary coordinates $x^\mu =(T,\lambda,\vartheta,\varphi)$, adapted to both the discrete self-similarity and axisymmetry, for diagnostic purposes.  In these coordinates $g_{\mu\nu}=e^{-2T}\bar g_{\mu\nu}(T,\lambda,\vartheta)$ with $\bar g_{\mu\nu}(T+\Delta,\lambda,\vartheta)=\bar g_{\mu\nu}(T,\lambda,\vartheta)$, meaning that $\Phi$ manifests itself as a periodicity in $T$ with echoing period $\Delta$.  There are many such coordinate systems in general.  Here, let $\tau$ be the proper time of an observer at the center, and $\tau_*$ the accumulation point of the self-similarity.  We then define $T\equiv -\ln(\tau_*-\tau) + T_0$, where both $\tau_*$ and $T_0$ depend on the family of initial data.  We also define the lines of constant {$(T,\vartheta,\varphi)$ to be null geodesics, starting from the center in the direction $(\vartheta = \theta,\varphi)$ at time $T$, and with affine parameter $\lambda$ normalized such that $\lambda = 0$ and $(dt/d\lambda)_T (dT/dt)_r = 1$ at the center.

\begin{figure*}
\includegraphics[width = 0.48 \textwidth]{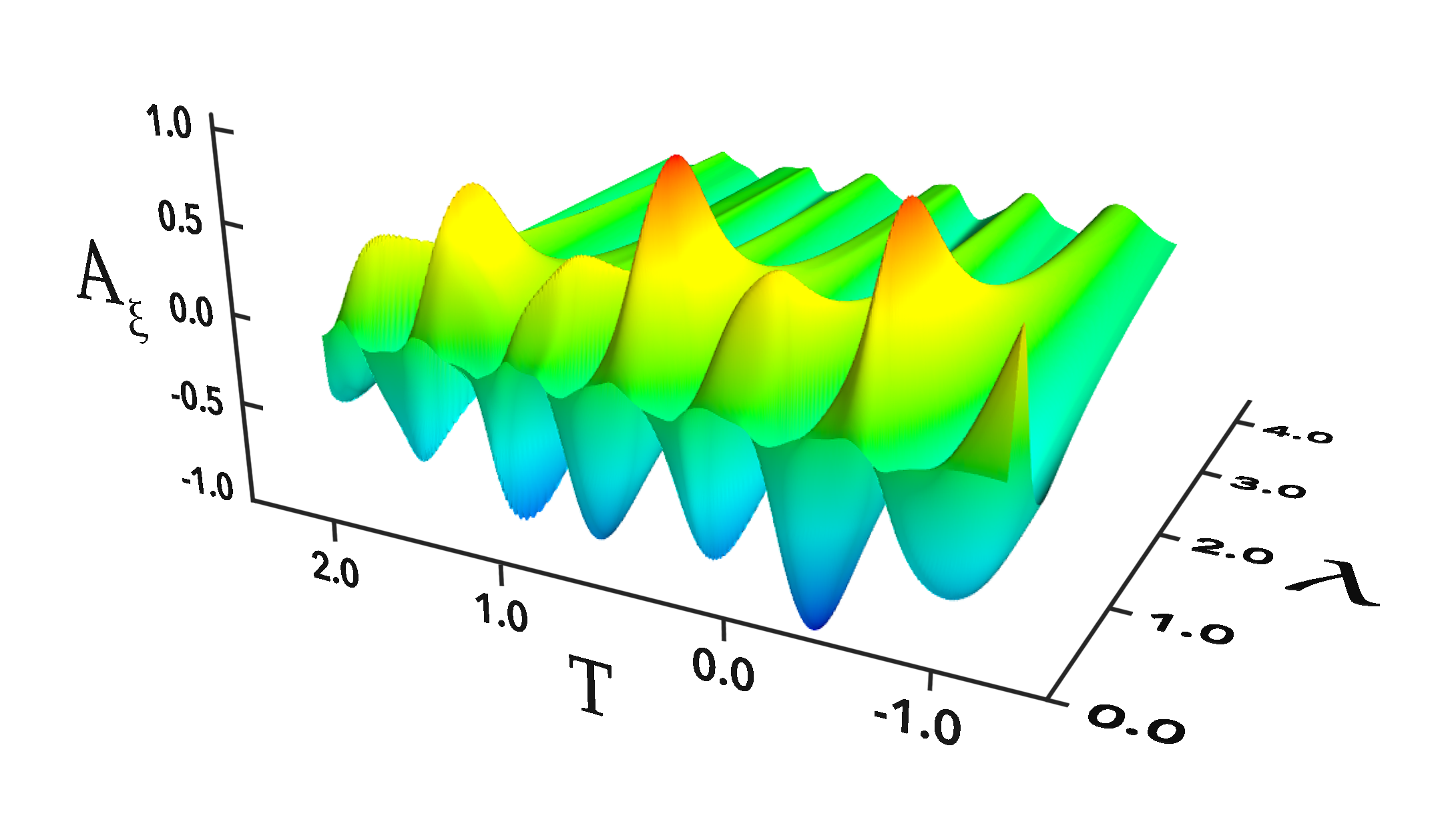}
\includegraphics[width = 0.48 \textwidth]{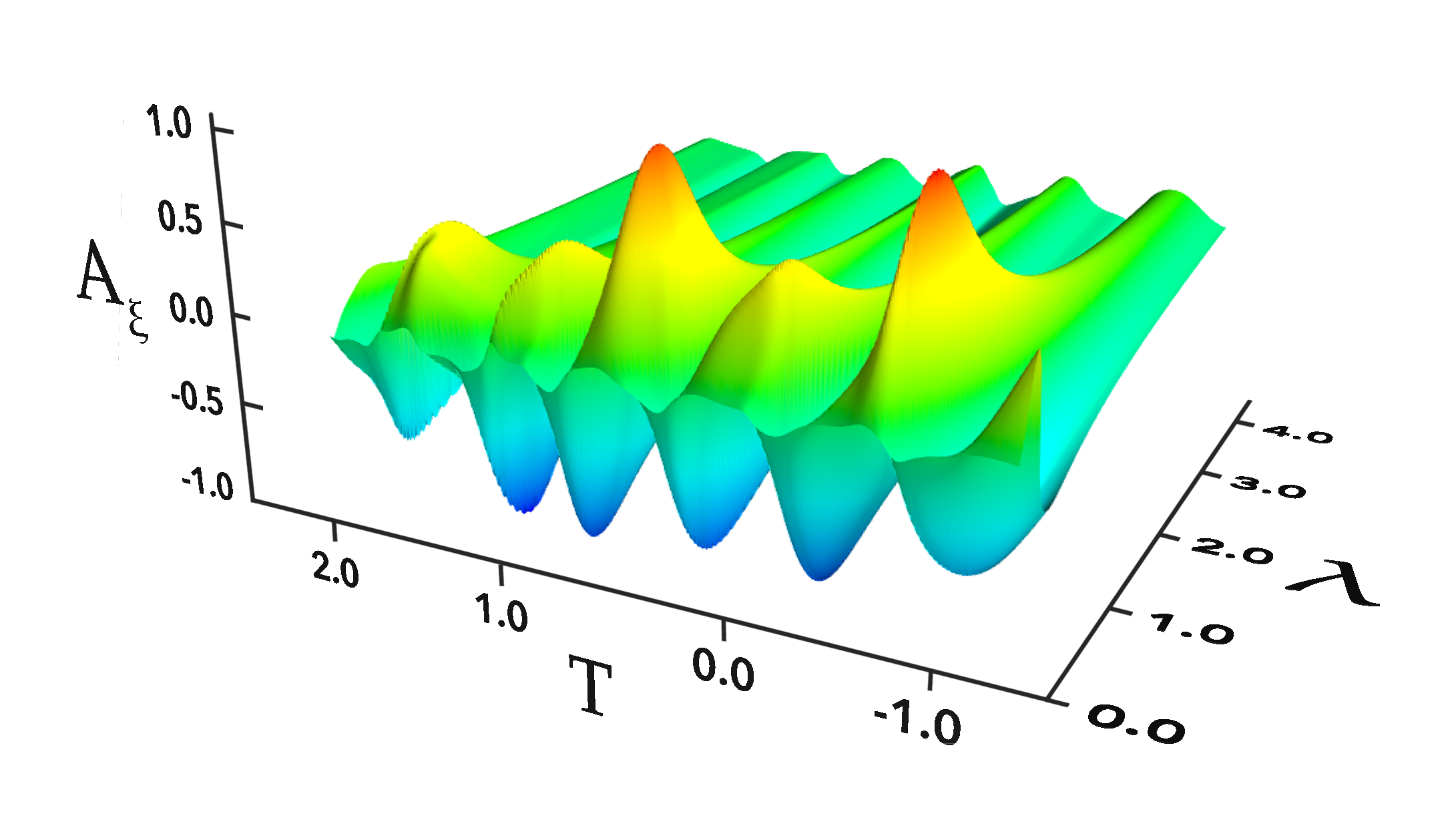}
\caption{Plots of $A_{\xi}$ in the equatorial plane ($\theta = \pi / 2$) as a function of the affine parameter $\lambda$ of null geodesics that originate from the center at time $T$.  The left panel shows results for centered data ($r_0 = 0$ with $\tau_* = 5.66$ and $T_0 = 0$), while the right panel shows results for off-centered data ($r_0 = 3$ with $\tau_* = 10.58$ and $T_0 = 0.42$).  }
\label{Fig:A_xi}
\end{figure*}

In Fig.~\ref{Fig:A_xi} we show plots of $A_\xi$ as a function of $\lambda$ and $T$ for near-threshold solutions in both the centered and the off-centered families.  While the graphs are not identical, they show remarkable similarities at intermediate times $-1.5 \lesssim T \lesssim 2.0$.   We take these similarities as an indication of at least an approximate universality of this threshold solution.  

It is also evident from Fig.~\ref{Fig:A_xi}, however, that the threshold solution is not strictly periodic.  A Fourier analysis of $A_\xi$ along lines of constant $\lambda$ shows a peak frequency that corresponds to an echoing period of $\Delta \simeq 0.55$ for both the centered and off-centered data.  

\begin{figure}
\includegraphics[width = 0.48 \textwidth]{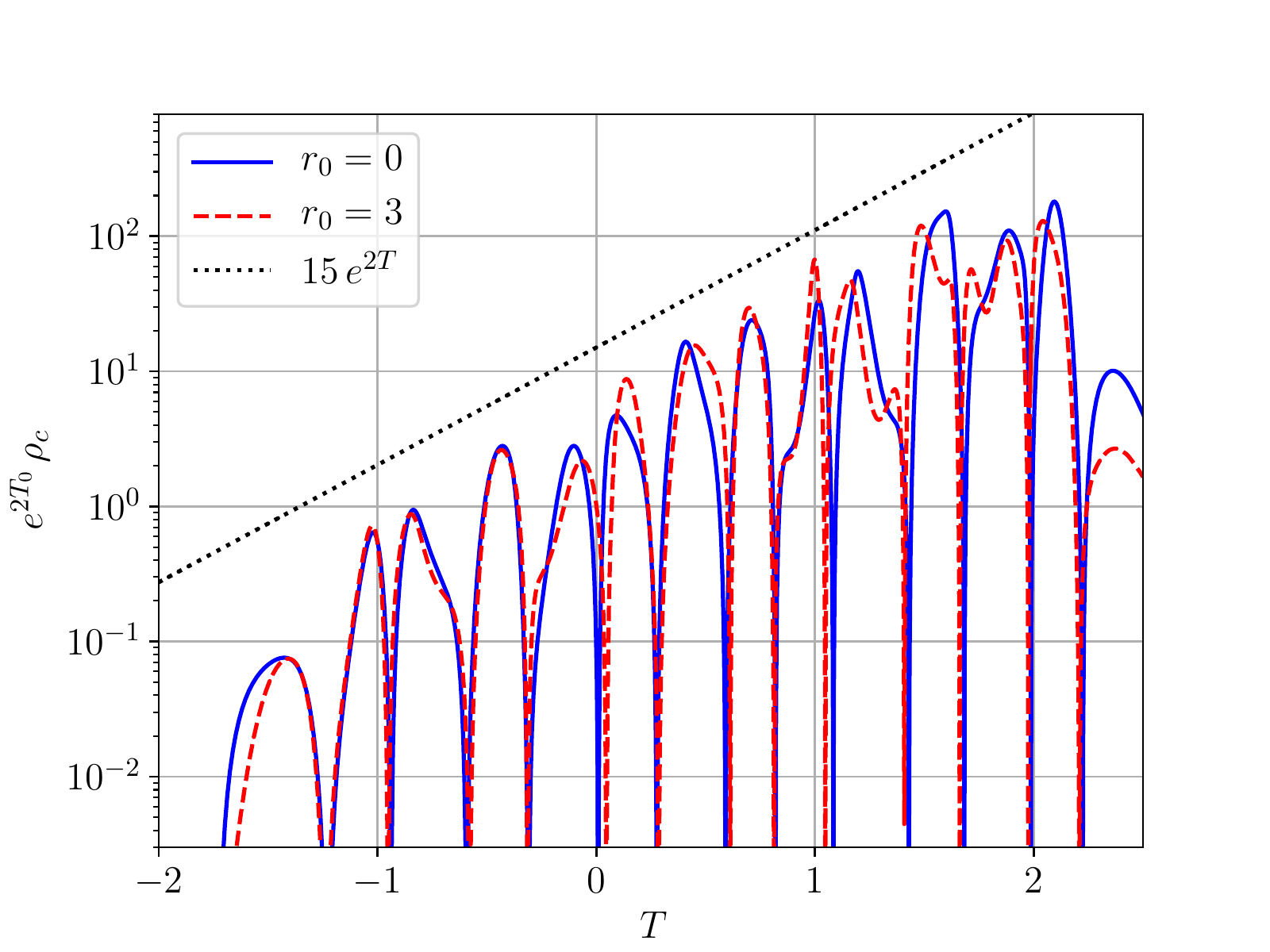}
\caption{The density $\rho$, evaluated at the center, as a function of $T = -\ln(\tau_* - \tau) + T_0$ for near-critical centered and off-centered evolutions.  In both cases the amplitude of the central density's oscillation increases approximately with $(\tau_* - \tau)^{-2} = e^{2(T - T_0)}$, which is consistent with self-similar contraction.  Both evolutions also display similar features, again suggesting an approximate universality.  As before, however, the oscillations are not strictly periodic, indicating that the critical solution is not exactly DSS.}
\label{Fig:rho_center}
\end{figure}

The absence of a strict periodicity is also visible in Fig.~\ref{Fig:rho_center}, where we show the density (\ref{density}) evaluated at the center as a function of $T$ for near-critical centered and off-centered evolutions.  The amplitude of the central density's oscillations increase approximately as expected for self-similar contraction, and both evolutions display similar features, hinting at some notion of universality in the critical solution -- but again the oscillations are not strictly periodic, suggesting that the critical solution is not exactly DSS.

\begin{figure}
\includegraphics[width = 0.48 \textwidth]{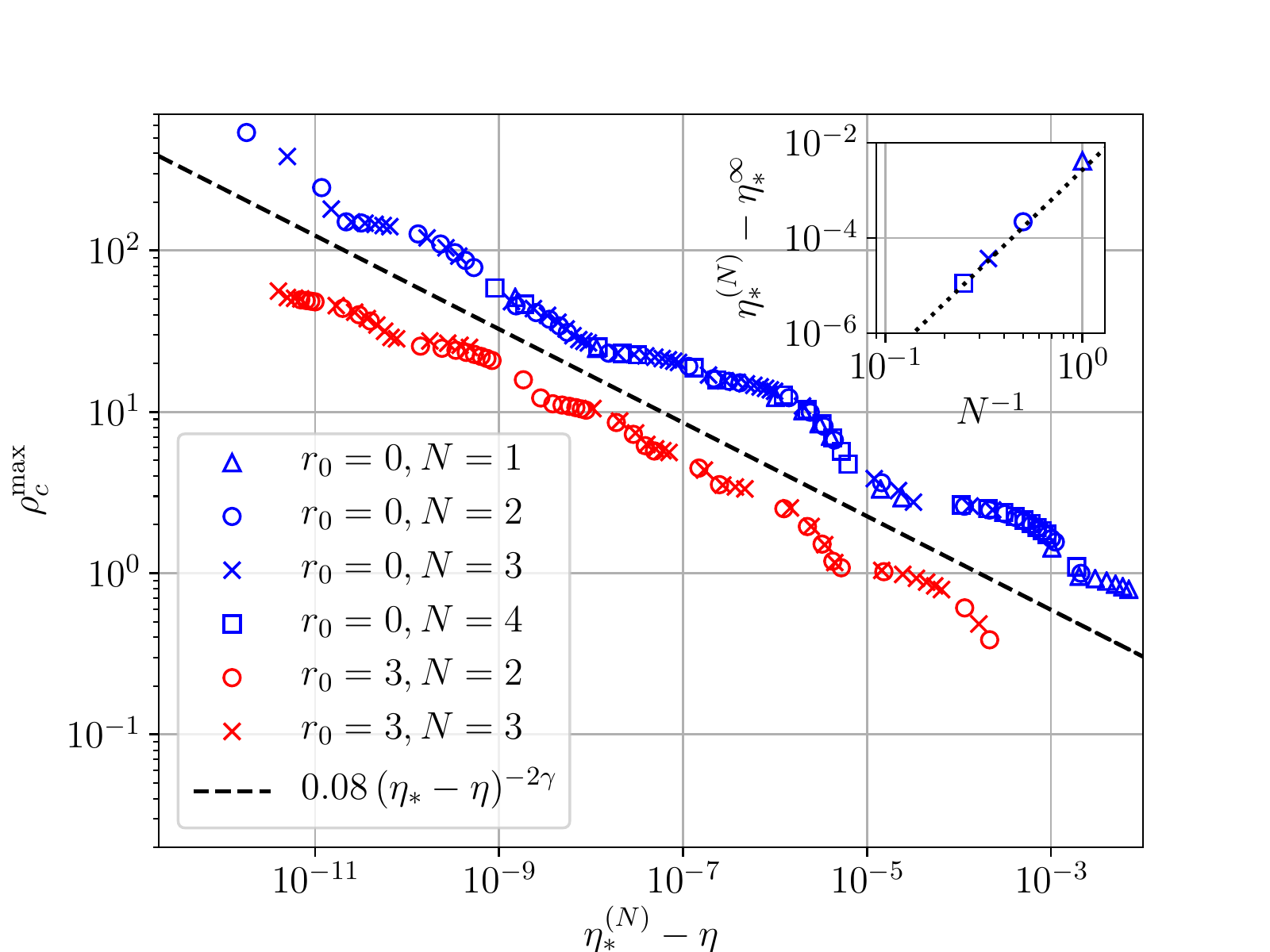}
\caption{The maximum central density for subcritical centered ($r_0 = 0$, blue, above the black dashed line) and off-centered ($r_0 = 3$, red, below the dashed line) evolutions, using $N_r = 64 N$ radial and $N_\theta = 6 N$ angular grid points.  The dashed line corresponds to scaling with $\gamma = 0.145$.  The fitted values of $\eta_*^{(N)}$ depend on the resolution $N$. Convergence of these values for $r_0 = 0$ is demonstrated in the inset, where we have adopted a Richardson extrapolated value of $\eta_*^\infty = 0.912895$, and where the dotted line is proportional to $N^{-4}$, indicating the expected fourth-order convergence.}
\label{Fig:rho_scaling}
\end{figure}

The power-law scalings (\ref{mass_scaling}) and (\ref{rho_scaling}) are a result of the growth of linear perturbations of the critical solution.  Different fine-tuning, i.e.~different values of $|\eta_* - \eta|$, lead to different size perturbations, which therefore become non-linear at different times.  The length scale of the self-similar solution at this moment endows the subsequent evolution with a length scale, and hence determines dimensional quantities like the black-hole mass and the maximum central density.  For CSS critical solutions, the power laws (\ref{mass_scaling}) and (\ref{rho_scaling}) are exact, while the periodicity of a DSS critical solution results in a periodic ``wiggle" that is superimposed on the scaling laws (see \cite{Gun97,HodP97}).  Given that we do not find an exactly DSS critical solution, we also do not expect deviations from power-law scalings to be exactly periodic.   This can be seen in Fig.~\ref{Fig:rho_scaling}, where we plot the maximum encountered central density as a function of $\eta_* = \eta$, where we have fit the values of $\eta_*$ to obtain behavior as close to power laws as possible.  While our results for both centered and off-centered evolutions approximately follow a power law (\ref{rho_scaling}) with $\gamma \simeq 0.145$ over several orders of magnitude, deviations from these power laws do not appear to be strictly periodic as one would expect for a strictly DSS critical solution.

In summary, our simulations suggest the existence of a self-similar critical solution at the threshold of black-hole formation in the gravitational collapse of electromagnetic waves.   Unlike in many other examples of critical collapse, however, the critical solution appears to be neither CSS nor exactly DSS; instead we observe only approximately periodic behavior.  Similarities between the critical solution obtained from different families of initial data hint at features of universality, but again this universality is not exact.  The absence of an exactly DSS critical solution is also reflected in the scaling behavior, which shows only approximately periodic deviations from a power law.

We suspect that this behavior is related to the absence of spherical symmetry.  For scalar fields and fluids, for example, the critical solution is spherically symmetric.  In this case, the critical solution can be described by spherical modes alone, and the gravitational fields do not possess independent degrees of freedom.   For electromagnetic waves, however, the critical solution cannot be spherically symmetric.  Given the nonlinear nature of the critical solution, it therefore cannot be described by just one angular mode; moreover, the gravitational fields can now carry gravitational radiation and hence possess independent degrees of freedom.

One possible explanation therefore invokes the competition between the critical solution of the electromagnetic waves and that of the gravitational waves.  As a toy model for critical phenomena in the collapse of two competing dynamical systems we recently studied the mixture of scalar fields and Yang-Mills fields, both restricted to spherical symmetry \cite{GunBH19}.  We found that, at sufficiently late times with sufficiently good fine-tuning, the scalar field always dominates.  At earlier times, however, the critical solution may be described as a mixture of both fields' critical solutions, and the transition from the dominance of one field to the other can be observed, for example, in the scaling laws.  While our findings for electromagnetic fields do not suggest such a transition from the dominance of one system to another, it would be of interest to generalize our work and evolve electromagnetic-wave initial data with different initial gravitational wave content.  This could be done, for instance, by choosing the initial conformally related metric to represent a gravitational wave, rather than flat space.

It is also possible, however, that the absence of a strictly DSS critical solution is inherently related to the presence of multiple angular modes.  One might attempt to describe such a system in terms of ``multi-oscillators" akin to those of \cite{ChoMW19}.   The presence of different  non-commensurate frequencies could explain the absence of an exact periodicity; moreover, for different families of initial data different ``oscillators" might enter with different phases, possibly explaining the absence of an exact universality.  

While we can only speculate about what causes the absence of an exactly DSS critical solution in the collapse of electromagnetic waves, our findings may well have bearing on critical phenomena in the collapse of vacuum gravitational waves, for which the critical solution also cannot be spherically symmetric.  While this critical solution is often expected to be DSS, we are not aware of any firm evidence -- either analytical or numerical -- that supports this hypothesis.   Still to this date, the strongest evidence was presented by Abrahams and Evans \cite{AbrE93,AbrE94}.  In Fig.~6 of \cite{AbrE94}, for example, they show approximate echoing in metric functions.  While the echoes do not overlap exactly (note also the absence of a periodicity in their Fig.~2), Abrahams and Evans attributed the differences to uncertainties in the determination of the echoing period, as well as the lack of sufficient fine-tuning.  In light of our findings here, however, we wonder whether their results instead provided the first suggestion that the critical solution in the collapse of vacuum gravitational waves is, indeed, not exactly DSS.  Similarly, \cite{HilWB17} found deviations from a simple power-law in the scaling of the Kretschmann scalar for the collapse of vacuum gravitational waves (see their Fig.~4), but could not establish these deviations to be periodic.   Comparing with our Fig.~\ref{Fig:rho_scaling}, we again suspect that these deviations are indeed not periodic, and instead evidence of the absence of an exact DSS in the critical solution for vacuum gravitational waves. 

\acknowledgements

This research was supported through the program Research in Pairs by the Mathematisches Forschungsinstitut Oberwolfach in 2019.  It is a pleasure to thank the institute and its staff for hospitality during our stay.  This work was also supported in parts by NSF grant PHYS-1707526 to Bowdoin College, as well as through sabbatical support from the Simons Foundation (Grant No.~561147 to TWB).  DH is supported by the FCT (Portugal) IF Program IF/00577/2015 and PD/BD/135511/2018.  The authors would also like to acknowledge networking support by the COST Action GWverse CA16104.  Numerical simulations were performed on the Bowdoin Computational Grid. We would like to thank Paul Howell for help with Fig.~\ref{Fig:A_xi}, which was produced using {\tt mayavi} software.

%


\end{document}